\documentstyle{ioplppt}
\begin{document}

\title{Brane-world generalizations of the Einstein static universe}

\author{L\'{a}szl\'{o} \'{A} Gergely\dag\, and Roy Maartens\ddag}

\address{\dag Astronomical Observatory and Department of
Experimental Physics, University of Szeged, Szeged~6723, D\'{o}m
t\'{e}r~9, Hungary}

\address{\ddag Relativity and Cosmology Group, School of
Computer Science and Mathematics, Portsmouth University,
Portsmouth~PO1~2EG, Britain}

\begin{abstract}

A static Friedmann brane in a 5-dimensional bulk (Randall-Sundrum
type scenario) can have a very different relation between the
density, pressure, curvature and cosmological constant than in the
case of the general relativistic Einstein static universe. In
particular, static Friedmann branes with zero cosmological
constant and 3-curvature, but satisfying $\rho>0$ and $\rho+3p>0$,
are possible. Furthermore, we find static Friedmann branes in a
bulk that satisfies the Einstein equations but is not
Schwarzschild-anti de Sitter or its specializations. In the models
with negative bulk cosmological constant, a positive brane tension
leads to negative density and 3-curvature.
\end{abstract}

\section{INTRODUCTION}

At high enough energies, Einstein's theory of general relativity
breaks down and is likely to be a limit of a more general theory.
In string theory/ M theory, gravity is a truly higher-dimensional
theory, becoming effectively 4-dimensional at lower energies.
Recent developments may offer a promising road towards a quantum
gravity theory~\cite{gen}. In brane-world models inspired by
string/M theory, the standard-model fields are confined to a
3-brane, while the gravitational field can propagate in $3+d$
dimensions (the `bulk'). The $d$ extra dimensions need not all be
small, or even compact: recently Randall and Sundrum~\cite{RS}
have shown that for $d=1$, gravity can be localized on a single
3-brane even when the fifth dimension is infinite. This noncompact
localization arises via the exponential `warp' factor in the
non-factorizable metric:
 \begin{equation}\label{a}
d\widetilde{s}\,^2=\exp(-2|y|/\ell)\left[-dt^2+d\vec{x}\,^2\right]
+dy^2\,.
 \end{equation}
For $y\neq0$, this metric satisfies the 5-dimensional Einstein
equations with negative 5-dimensional cosmological constant,
$\widetilde{\Lambda}\propto -\ell^{-2}$. The brane is located at
$y=0$, and the induced metric on the brane is a Minkowski metric.
The bulk is a 5-dimensional anti-de Sitter metric, with $y=0$ as
boundary, so that $y<0$ is identified with $y>0$, reflecting the
$Z_2$ symmetry, with the brane as fixed point, that arises in
string theory.

Perturbation of the metric~(\ref{a}) shows that the Newtonian
gravitational potential on the brane is recovered at lowest order:
 \begin{equation}\label{newt}
V(r) = {GM\over r}\left(1+{2\ell^2\over 3r^2}\right)+\cdots
 \end{equation}
Thus 4-dimensional gravity is recovered at low energies, with a
first-order correction that is constrained by current
sub-millimetre experiments. The lowest order term corresponds to
the massless graviton mode, bound to the brane, while the
corrections arise from massive Kaluza-Klein modes in the bulk.
Generalizing the Randall-Sundrum model to allow for matter on the
brane leads to a generalization of the metric~(\ref{a}), and in
general to a breaking of conformal flatness, since matter on the
brane in general induces Weyl curvature in the bulk. Indeed, the
massive Kaluza-Klein modes that produce the corrective terms in
Eq.~(\ref{newt}) reflect the bulk Weyl curvature that arises from
a matter source on the brane.

The 5-dimensional Einstein equations with a brane at $y=0$
containing a general energy-momentum tensor are
\begin{equation}
\widetilde{G}_{AB} =
\widetilde{\kappa}^2\left[-\widetilde{\Lambda}\widetilde{g}_{AB}
+\delta(y)\left\{ -\lambda g_{AB}+T_{AB}\right\}\right]\,,
\label{1}
\end{equation}
where $\widetilde{\kappa}^2= 8\pi/\widetilde{M}_{\rm p}^3$, with
$\widetilde{M}_{\rm p}$ the fundamental 5-dimensional Planck mass,
which is typically much less than the effective Planck mass on the
brane, $M_{\rm p}=1.2\times 10^{19}$ GeV. The brane tension is
$\lambda$, and standard-model fields confined to the brane make up
the brane energy-momentum tensor $T_{AB}$, with $T_{AB}n^B=0$,
where $n^A=\delta^A{}_y$ is the unit normal to the brane. Using
the Gauss-Codazzi equations, the Darmois-Israel matching
conditions and the $Z_2$ symmetry about the brane, one can derive
the induced field equations on the brane~\cite{SMS}:
\begin{equation}
G_{\mu\nu}=-\Lambda g_{\mu\nu}+\kappa^2
T_{\mu\nu}+\widetilde{\kappa}^4S_{\mu\nu} - {\cal E}_{\mu\nu}\,,
\label{2}
\end{equation}
where $\kappa^2=8\pi/M_{\rm p}^2$ and
$g_{AB}=\widetilde{g}_{AB}-n_An_B$. The energy scales are related
to each other via
\begin{equation}
\lambda=6{\kappa^2\over\widetilde\kappa^4} \,, ~~ \Lambda
={\textstyle{1\over2}}\widetilde\kappa^2
\left(\widetilde{\Lambda}+{\textstyle{1\over6}}\widetilde\kappa^2
\lambda^2\right)\,. \label{3}
\end{equation}
The higher-dimensional modifications of the standard Einstein
equations on the brane are of two forms: first, the matter fields
contribute local quadratic energy-momentum corrections via the
tensor $S_{\mu\nu}$, which arise from the extrinsic curvature, and
second, there are nonlocal effects from the free gravitational
field in the bulk, transmitted via the projection onto the brane
of the bulk Weyl tensor: ${\cal
E}_{AB}=\widetilde{C}_{ACBD}n^Cn^D$, and ${\cal E}_{\mu\nu}$ on
the brane is given by the limit as $y\to0$. The local corrections
are given by
\begin{equation}
S_{\mu\nu}={\textstyle{1\over12}}T_\alpha{}^\alpha T_{\mu\nu}
-{\textstyle{1\over4}}T_{\mu\alpha}T^\alpha{}_\nu+
{\textstyle{1\over24}}g_{\mu\nu} \left[3 T_{\alpha\beta}
T^{\alpha\beta}-\left(T_\alpha{}^\alpha\right)^2 \right]\,.
\label{3'}
\end{equation}

\section{STATIC FRIEDMANN BRANES}

The generalized Friedmann equation on a spatially homogeneous and
isotropic brane is~\cite{BDEL}
\begin{equation}\label{f}
H^2={\textstyle{1\over3}}\kappa^2\rho\left(1+{\rho\over
2\lambda}\right)+{\textstyle{1\over3}}\Lambda -{K\over a^2} +
{2{\cal U}_o\over\kappa^2\lambda} \left({a_o\over a}\right)^4\,,
\end{equation}
where $K=0,\pm1$. The $\rho^2/\lambda$ term is the $S_{\mu\nu}$
contribution, which is significant only at high
energies~\cite{mwbh}: $\rho>\lambda>$ (100 GeV)$^4$. The ${\cal
U}_o/a^4$ term is the ${\cal E}_{\mu\nu}$ contribution. General
relativity is recovered in the limit $\lambda^{-1}\to0$. The
generalized Raychaudhuri equation~\cite{Maartens} becomes, for a
Friedmann brane,
 \begin{equation}\label{r}
\dot H+H^2=-{\textstyle{1\over6}} \kappa^2\left[\rho+3p+(2\rho+3p)
{\rho\over \lambda}\right]+{\textstyle{1\over3}}\Lambda -{2{\cal
U}_o\over\kappa^2\lambda} \left({a_o\over a}\right)^4\,.
 \end{equation}

$Z_2$-symmetric Friedmann branes can be embedded in 5-dimensional
Schwarzschild-anti de Sitter (SAdS) space~\cite{MSM,BCG}, with the
mass parameter of the bulk black hole proportional to ${\cal
U}_o$. These solutions include the special case of static
Friedmann branes. As we show in the next section, there are other
bulk solutions, which are not SAdS, but which do admit a
$Z_2$-symmetric static Friedmann brane.

For a static Friedmann brane, we have $a=a_o$ and $H=0$, so that
Eqs.~(\ref{f}) and (\ref{r}) become
\begin{eqnarray}
&&\kappa ^{2}\rho\left( 2 +\frac{\rho }{\lambda }\right) =6{K\over
a_o^2}- \frac{ 12{\cal U}_o}{\kappa ^{2}\lambda }-2\Lambda
\,,\label{fs} \\ &&\kappa ^{2}\left[ 3p+\rho +{\rho\over\lambda}
(3p+2\rho )\right] =- \frac{ 12{\cal U}_o}{\kappa ^{2}\lambda
}+2\Lambda \,.\label{rs}
\end{eqnarray}
The general relativity Einstein static universe is the case
$\lambda^{-1}=0$, which leads to
\begin{eqnarray}
\kappa^2\rho = 3{K\over a_o^2}-\Lambda\,,~~ \kappa^2p = -{K\over
a_o^2}+\Lambda\,,\label{rse}
\end{eqnarray}
so that if $\rho>0$ and $\rho+3p>0$, then $K=+1$ and $\Lambda>0$.

For the static brane-world model we find
\begin{eqnarray}
&&\kappa^2\rho\left(1+{\rho\over2\lambda}\right) = 3{K\over
a_o^2}-\Lambda-{6{\cal U}_o\over\kappa^2\lambda}\,,\label{fsb}\\
&& \kappa^2\left[p+{\rho\over2\lambda}(2p+\rho)\right] = -{K\over
a_o^2}+\Lambda-{2{\cal U}_o\over\kappa^2\lambda}\,.\label{rsb}
\end{eqnarray}
The local (i.e., $\rho/\lambda$) and nonlocal (i.e., ${\cal
U}_o/\lambda$) terms introduce new possibilities compared with
general relativity. If  $\rho>0$ and $\rho+3p>0$, then
\begin{equation}
\Lambda>{6{\cal U}_o\over\kappa^2\lambda}\,,~~~~3{K\over a_o^2}
-\Lambda> {6{\cal U}_o\over\kappa^2\lambda}\,.
\end{equation}
Thus we can satisfy $\rho>0$ and $\rho+3p>0$ with $K=0=\Lambda$,
provided that ${\cal U}_o<0$ and
\begin{equation}
\rho-3p={\rho\over\lambda}(\rho+3p)\,.
\end{equation}
This equation has no general relativity limit, since ${\cal
U}_o<0$, and it implies $p<{1\over3}\rho$.

The general static solution satisfies equations~(\ref{fsb}) and
(\ref{rsb}). This includes the solutions with $p=(\gamma-1)\rho$,
$\gamma$ constant, given in~\cite{cs}, which are saddle points in
the dynamical phase space (like the Einstein static solution in
general relativity).

\section{NON-SAdS BULK WITH STATIC FRIEDMANN BRANE}

In the 5-dimensional Einstein equation~(\ref{1}), we write
\begin{equation}
\widetilde{\kappa}^2\widetilde{\Lambda}=3\epsilon\Gamma^2\,,
~\Gamma>0\,,~\epsilon=0,\pm 1\,, \label{Ein5}
\end{equation}
where $\Gamma$ gives the magnitude of the cosmological constant
and $\epsilon$ its sign; if $\epsilon=0$, then $\Gamma$ is a
removable constant. Then we find the following solution of
Eq.~(\ref{1}) for $y>0$:
\begin{eqnarray}
\Gamma^2 d\,\widetilde{s}\,^{2} &=&-F^{2}(y;\epsilon)dt^{2}+d\chi
^{2}+{\cal H}^{2}\left( \chi ;\epsilon\right) \left( d\theta
^{2}+\sin ^{2}\theta d\varphi ^{2}\right) +dy^{2}\,,
\label{metric5}
\end{eqnarray}
where
\begin{eqnarray}
 F(y;\epsilon) &=&\left\{
\begin{array}{cc}
A\cos \ \left( \sqrt{2}\ y\right) +B\sin \ \left( \sqrt{2}\
y\right)  & ,\qquad \epsilon=1\,, \\ A+\sqrt{2}\ By\  & ,\qquad
\epsilon=0\,, \\ A\cosh \ \left( \sqrt{2}\ y\right) +B\sinh \
\left( \sqrt{2}\ y\right)  & \ ,\qquad \epsilon=-1\,,
\end{array}
\right.   \nonumber \\ {\cal H}(\chi ;\epsilon) &=&\left\{
\begin{array}{cc}
\sin \ \chi \  & ,\qquad \epsilon=1\,, \\ \chi  & ,\qquad
\epsilon=0\,,
\\ \sinh \ \chi  & \ ,\qquad \epsilon=-1\,.
\end{array}
\right. \   \nonumber
\end{eqnarray}
Either of the constants $A$ or $B$ can be absorbed into the
coordinate $t$, so that we have a one-parameter family of
solutions. The derivatives of the metric functions obey ($y>0$)
\begin{eqnarray}
\left( \partial _{y}F\right) ^{2} &=&\alpha\beta ^{2}-2\epsilon
F^{2}\ ,\qquad
\partial _{y}^{2}F=-2\epsilon F \ , \nonumber \\ \left( \partial _{\chi
}{\cal H}\right) ^{2} &=&1-\epsilon {\cal H}^{2}\ ,\qquad \qquad
\partial _{\chi }^{2}{\cal H}=-\epsilon {\cal H} \,,
\end{eqnarray}
where $\alpha={\rm sgn}\,(\epsilon A^{2}+B^{2})$ and $\beta
=\sqrt{2 |\epsilon A^{2}+B^{2}| }$. Note that $\alpha=0,-1$ can
occur for $\epsilon =-1$. We exclude the case $B=0$, since, as is
shown below, the vanishing of $B$ is incompatible with matter on
the brane ($y=0$).

Other useful formulas for the above functions will be also
employed later:
\begin{equation}
\partial_{y}\log F+\alpha\beta^{2}\int \frac{dy}{F^{2}}
= -\sqrt{2}\frac{A}{B}\epsilon  \,,~~
\partial _{\chi }\log {\cal H}+\int \frac{d\chi }{{\cal H}^{2}} =0\,.
\end{equation}

The projected part of the bulk Weyl tensor is
\begin{equation}
{\cal E}_{AB}=\tilde{C}_{ACBD}n^{C}n^{D}=-{\textstyle{3\over2}}
\epsilon\Gamma^2 \left( u_{A}u_{B}+{\textstyle{1\over3}}
h_{AB}\right)\,, \label{Weylproj}
\end{equation}
where $u^A=\Gamma F^{-1}\delta^A{}_0$ is the 4-velocity along the
static Killing vector of the metric~(\ref{metric5}), and
$h_{AB}=g_{AB}+u_A u_B$. The general solution of the Killing
equation is given in the appendix.

The ${Z}_{2}$-symmetric solution is given by Eq.~(\ref{metric5})
with $y$ replaced by $|y|$. On the brane $y=0$, the induced metric
is
\begin{equation}
\Gamma^2 d{s}^{2} =-A^{2}dt^{2}+d\chi ^{2}+{\cal H}^{2}\left( \chi
;\epsilon\right) \left( d\theta ^{2}+\sin ^{2}\theta d\varphi
^{2}\right)\,, \label{metric4}
\end{equation}
which has Friedmann (with curvature index $\epsilon$) and static
symmetry. The vectors $K_{{\bf 1-7}}$ in Eq.~(\ref{KilVec}) are
the Killing vectors for this metric. The scale factor is
\begin{equation}\label{radius}
a_o={1\over\Gamma}\,.
\end{equation}

By Eq.~(\ref{3}), the brane cosmological constant $\Lambda $ is
related to the brane tension and bulk cosmological constant via
\begin{equation}
\Lambda ={\textstyle{1\over2}}\left(3\epsilon\Gamma^2+\kappa
^{2}\lambda \right) \,. \label{lambdas}
\end{equation}
In the space-time given locally by the metric (\ref{metric5}), the
extrinsic curvature of any hypersurface $y=$const. has only one
nonvanishing component
\begin{equation}
K_{00}=-{1\over\Gamma} F\left( y\right) \partial _{y}F\left(
y\right) \,, \label{K}
\end{equation}
so that the jump in the extrinsic curvature across the brane is
\begin{equation}
[K_{\mu\nu}]^+_-=-{2\sqrt{2}B\over \Gamma}\, \delta_{\mu}{}^{0}
\delta_{\nu}{}^{0}\,.
\end{equation}
Since $[K_{\mu\nu}]^+_-\propto
T_{\mu\nu}+{1\over3}(\lambda-T)g_{\mu\nu}$, it follows that the
brane energy density is
\begin{equation}\label{23}
\rho=-\lambda\,.
\end{equation}
Since the low energy limit of the modified Einstein equations
implies $\lambda >0$, our bulk solution has {\em negative} energy
density on the brane, irrespective of the sign of the bulk
cosmological constant.

\section{COMPARISON WITH RELATED SOLUTIONS}

Several recent works have tackled the issue of Friedmann-type
branes embedded in 5-dimensional bulk spacetimes, and we can
compare our solution with these to see if ours is simply a special
case.

Equations (2), (23), (37) and (38) of~\cite{BDEL} describe a bulk
solution in Gaussian normal coordinates adapted to the Friedmann
brane which is at $y=0$. However in the limit of a static brane,
their bulk metric function $n\left( t,y\right) $ remains
undefined.

The most general bulk solution in which a static, maximally
symmetric 3-space is embedded, is claimed to be found
in~\cite{Mannheim}. However the line element given by Eqs. (3) and
(4) of \cite{Mannheim} does {\em not} include our solution,
Eq.~(\ref{metric5}). This can be seen as follows. After the
transformation $r={\cal H}(\chi ;k)$, we see that the attempt to
make the two line elements coincide requires that $k=\epsilon=-1$
and that the metric function $f$ of \cite{Mannheim} must be a
constant, $f=\Gamma^{-2}$. The latter can be achieved for the
values $\alpha =\beta =0$ of the parameters in~\cite{Mannheim}.
But for these parameters the other metric function $\ e$ is
vanishing, and in consequence the bulk metric is singular.

In \cite{MSM}, the equivalence is proved between the bulk solution
found in~\cite{BDEL} and the SAdS bulk, given by
\begin{eqnarray}
&& d\,\widetilde{s}\,^{2} =
-f(r;K)dt^{2}+\frac{dr^{2}}{f(r;K)}+r^{2}\left[ d\chi ^{2}+{\cal
H}^{2}\left( \chi ;K\right) \left( d\theta ^{2}+\sin ^{2}\theta
d\varphi ^{2}\right) \right] ,  \nonumber \\&& f(r;K) =
K+{\Gamma^2\over2} r^{2}-\frac{\mu }{r^{2}}\,, \label{Sch-AdS5}
\end{eqnarray}
where the brane is described by a moving domain wall (see also
\cite{Kraus}). Here $K=0,\pm
1$ and $\mu$ is a constant which is proportional to ${\cal U}_o$:
\begin{equation}
{3\mu \over a^{4}}=-{\cal E}_{\mu\nu}u^{\mu}u^{\nu}\,.
\end{equation}
This relation follows by comparing the generalized Friedmann
equation on the moving brane in SAdS spacetime, Eq.~(3) of
\cite{MSM}, with the generic form of this equation, Eq.~(\ref{f}).
Then Eqs.~(\ref{Weylproj}) and (\ref{radius}) imply that our bulk
solution with negative cosmological constant is characterized by
\begin{equation}
\mu =-\frac{1}{2\Gamma^2}\,.  \label{mu}
\end{equation}
For this value of $\mu $ and vanishing Hubble constant, in the
hyperbolic case ($K=-1$) the metric functions of~\cite{MSM} become
$\phi =1$ and $\psi =\cosh (\sqrt{2}\Gamma w).$ By identifying
$y=\Gamma  w$, we recover the $\epsilon=-1$ case of our metric,
Eq.~(\ref{metric5}), but with the constant $B=0.$ Therefore our
generic bulk metric is {\em not} contained in the analysis
of~\cite{MSM}.

Alternatively, in the case $A>B$ we can identify our bulk solution
with the metric in Eq.~(23) of~\cite{MSM} by inserting $y=\Gamma
w-\eta $, where $\sqrt{2}\eta = \tanh^{-1}( B/A)$. However, the
brane is then confined to $w=\eta / \Gamma $ in our approach,
and to $w=0$ in~\cite{MSM}.

SAdS spacetime admits an event horizon given by $f(r;K)=0$:
\begin{equation}
r_{\rm h}^{2}=\frac{1}{\Gamma^2}\left( -K+\sqrt{K^{2}+2 \Gamma^2
\mu }\right) \,.
\end{equation}
We note that the coordinate transformation employed in~\cite{MSM}
is singular at $r_{\rm h}$, since the affine parameter $w$
diverges, so that it cannot be used as a new coordinate.  (In our
solution, $ r_{\rm h}=a_o$.) Thus we cannot repeat the procedure
of~\cite{MSM} in order to transform our bulk
metric~(\ref{metric5}) to the SAdS form~(\ref{Sch-AdS5}).

An elegant approach in~\cite {BCG} leads to the generic constant
curvature bulk solution containing a constant spatial curvature
brane, generalizing Taub's solution. By solving the 5-dimensional
Einstein equations for the {\it non-constant} metric functions
$B,\nu $, they find the generic solution in terms of derivatives
of $B$. By introducing the radial coordinate $r=B^{1/3}$, they
show that this is nothing other than the SAdS spacetime in 5
dimensions. The metric ansatz Eq.~(3) of~\cite{BCG}, found from
symmetry considerations, does include our solution,
Eq.~(\ref{metric5}), when their metric functions take the
particular values $B^{2/3}=\Gamma^{-2}=a_o^{2}$ and $e^ \nu
 =\Gamma^{-2}F(y)$, and their fifth coordinate
$z$ is related to ours by $dz=dy/F(y)$. But their final result
Eq.~(20), containing derivatives of $B$, again does {\it not}
contain our metric~(\ref{metric5}), as our metric is characterized
by a constant $B$. Neither can $r=B^{1/3}=a_o$ be introduced as a
new coordinate.

In some sense our solution~(\ref{metric5}) resembles the
Bertotti-Robinson solution encountered in the conventional
4-dimensional Einstein theory. There the radius of the 2-spheres
is also constant, so that it cannot be chosen as a new coordinate.
However the Bertotti-Robinson solution can be interpreted as
describing the infinite throat of the extreme Reissner-Nordstrom
black hole. It would be interesting to find whether some relation
between the metric~(\ref{metric5}) and the SAdS solution holds.

For this purpose first we note that the Killing vectors of the
SAdS metric~(\ref{Sch-AdS5}) are nothing but $K_{{\bf 1-7}}$ given
by Eq.~(\ref{KilVec}), with $\alpha=0$ and $ {\cal H}={\cal
H}(\chi ;K)$. Further, for $\epsilon=-1$ the following scalars
agree for our bulk metric~(\ref{metric5}) and the SAdS
metric~(\ref{Sch-AdS5}):
$\widetilde{R}=\widetilde{g}^{AB}\widetilde{R}_{AB},\
\widetilde{R}_{AB}\widetilde{R}^{AB}$. This, however happens for
{\em all} solutions of the 5-dimensional Einstein
equation~(\ref{1}), which implies $\widetilde{R}=10\epsilon
\Gamma^2$ and $\widetilde{R}_{AB}= 2\epsilon\Gamma^2
\widetilde{g}_{AB}.$ However, there are other curvature scalars
which are different, showing that our solution is {\em not} SAdS.
Two examples are
\begin{equation}\label{s1}
\widetilde{R}_{ABCD}\,\widetilde{R}^{ABCD}=\left\{
\begin{array}{cc}
28\Gamma^4\,, & \qquad \mbox{our solution}\,,\\
10\Gamma^4+72\mu^2/r^8\,, & \qquad \mbox{SAdS}\,,
\end{array} \right.
\end{equation}
and
\begin{equation}\label{s2}
\widetilde{C}_{ABCD}\,\widetilde{C}^{ABCD}=\left\{
\begin{array}{cc}
18\Gamma^4\,, & \qquad \mbox{our solution}\,, \\ 72\mu^2/r^8\,, &
\qquad \mbox{SAdS}\,,
\end{array} \right.
\end{equation}

Since the scalar ${\cal E}_{AB}u^{A}u^{B}$ coincides for the two
metrics when the condition $\mu =-1/(2\Gamma^2)$ holds, we find
that the scalars in Eqs.~(\ref{s1}) and (\ref{s2}) agree at the
particular radial coordinate value $r=a_o$. Of course in 5
dimensions there are 40 independent curvature scalars to compare,
but still we have a serious indication that our solution with
$\epsilon=-1$ is related to the event horizon of that particular
SAdS space-time, hyperbolic case ($K=-1$), which has the smallest
possible horizon area.

\section{CONCLUDING REMARKS}

We have shown in Sec.~2 how branes of static Friedmann type in
general allow many more possibilities than the special limiting
case of the general relativistic Einstein static universe. We have
also found, in Sec.~3, a family of bulk solutions of the
5-dimensional Einstein equations with high symmetry, admitting a
$Z_2$-symmetric brane of static Friedmann type. These solutions
are {\em not} Schwarzschild-anti de Sitter (SAdS), as confirmed by
calculating the bulk curvature scalars
$\widetilde{R}_{ABCD}\,\widetilde{R}^{ABCD}$ and
$\widetilde{C}_{ABCD}\,\widetilde{C}^{ABCD}$.

Up to now, the known $Z_2$-symmetric Friedman branes in
Randall-Sundrum type models have all been embedded in an SAdS
bulk. One could easily assume that all such Friedmann branes must
be embedded in SAdS spacetime. However, our new solutions show
that this is {\em not} true; there are static $Z_2$-symmetric
Friedmann branes that are embedded in the non-SAdS bulk given by
Eq.~(\ref{metric5}). This is the main significance of our new
solutions. These solutions have negative matter energy density,
$\rho$. However, the {\em total} energy density on the brane
is~\cite{SMS,BDEL} $\rho_{\rm brane}=\rho+\lambda$, and by
Eq.~(\ref{23}) this is zero: $\rho_{\rm brane}=0$. By
Eq.~(\ref{lambdas}), the effective cosmological constant on the
brane is
\begin{equation}
\Lambda={1\over2}\left[\kappa^2\lambda+\epsilon \widetilde\kappa
^2|\widetilde\Lambda| \right]\,,
\end{equation}
which is positive for $\epsilon\geq0$. In the case $\epsilon=-1$,
one can always choose the bulk cosmological constant
$\widetilde\Lambda$ so that $\Lambda$ is positive or zero. Unlike
the general relativity Einstein static universe, the brane-world
for $\epsilon=-1$ has {\em negative} curvature; it is kept static
by {\em negative} nonlocal energy density:
\begin{equation}
{6\,{\cal U}_o\over\kappa^2\lambda}=-{1\over2}\widetilde\kappa^2
|\widetilde\Lambda|\,.
\end{equation}

\[ \]
\ack

We thank Marco Bruni, Toni Campos and David Wands for useful
discussions. L\'{A}G is grateful for the financial support of the
organizers of the Cosmology Workshop, Paris 2000, where this work
was initiated. His research was supported by the Zolt\'{a}n
Magyary fellowship.

\[ \]
\[ \]

\appendix
\section{Killing vectors}

From the general solution of the Killing equation for the
metric~(\ref {metric5}), we find the independent Killing vectors.
In the coordinates $x^A=(x^\mu,y)$, they are ($y>0$):
\begin{eqnarray}
K_{{\bf 1}} &=&\left( 0,0,0,1,0\right) \ ,  \nonumber \\ K_{{\bf
2}} &=&\left( 0,0,-\cos \varphi ,\cot \theta \sin \varphi
,0\right) \ ,  \nonumber \\ K_{{\bf 3}} &=&\left( 0,0,\sin \varphi
,\cot \theta \cos \varphi ,0\right) \ ,  \nonumber \\ K_{{\bf 4}}
&=&\left( 0,-\cos \theta ,\partial _{\chi }\left( \log {\cal
H}\right) \sin \theta ,0,0\right) \ ,  \nonumber \\ K_{{\bf 5}}
&=&\left( 0,\sin \theta \sin \varphi ,\partial _{\chi }\left( \log
{\cal H}\right) \cos \theta \sin \varphi ,\partial _{\chi }\left(
\log {\cal H}\right) \frac{\cos \varphi }{\sin \theta },0\right)
\!\ \!\!, \nonumber
\\ K_{{\bf 6}} &=&\left( \!0,\sin \theta \cos \varphi ,\partial
_{\chi }\left( \log {\cal H}\right) \cos \theta \cos \varphi
,\!-\!\partial _{\chi }\left( \log {\cal H}\right) \frac{\sin
\varphi }{\sin \theta },0\!\right) \ \!\!,  \nonumber \\ K_{{\bf
7}} &=&[\alpha\beta +\delta _{\alpha}{}^{0}]^{-1}\left(
1,0,0,0,0\right) \ , \nonumber
\\ K_{{\bf 8}} &=&\frac{1}{\sqrt{2}}\left[ \alpha^{2}\beta +
\frac{A }{B}\delta_{\alpha}{}^{0}\right] \left( -\frac{
\delta_{\alpha}{}^{0}}{2\sqrt{2}F^{2}}-\partial _{y}\left( \log
F\right) \int Ldt,0,0,0,L\right) \ ,  \nonumber \\ K_{{\bf 9}}
&=&\frac{1}{\sqrt{2}}\left[ \alpha +
\frac{A}{B}\delta_{\alpha}{}^{0} \right] \left( -\partial
_{y}\left( \log F\right) L,0,0,0,\partial _{t}L\right) \ ,
\label{KilVec}
\end{eqnarray}
where we have introduced the function
\begin{equation}
L\left( t;\alpha\right) =\left\{
\begin{array}{cc}
\beta^{-1} \cosh \beta t\ \  & ,\qquad \alpha=1\,, \\ t & ,\qquad
\alpha=0\,,
\\ \beta^{-1} \cos \beta t & \ ,\qquad \alpha=-1\,,
\end{array}
\right.
\end{equation}
obeying
\begin{equation}
\left( \partial _{t}L\right) ^{2}=1+\alpha\beta ^{2}L^{2}\ ,\qquad
\qquad
\partial _{t}^{2}L=\alpha\beta ^{2}L\ ,
\end{equation}
and we have used
\begin{equation}
\int \frac{dy}{F^{2}}=-\frac{1}{2\sqrt{2}F^{2}}
\end{equation}
for $\alpha=0$.

Among the Killing vectors $K_{{\bf 1-6}}$ are obviously spacelike
and $K_{{\bf 7}}$ is timelike, while for $K_{{\bf 8,9}}$ we find
that
\begin{eqnarray}
\widetilde{g}\left( K_{{\bf 8}},K_{{\bf 8}}\right)
&=&\frac{1}{2\Gamma^2}\left\{
\begin{array}{cc}
2\epsilon (F/\beta)^{2}[ 1+\alpha\beta ^{2}L^{2}] -\alpha\,, &
\qquad \alpha\neq 0\,, \\ -E^{2}<0\,, & \qquad \alpha=0\,,
\end{array}
\right.  \nonumber \\ \widetilde{g}\left( K_{{\bf 9}},K_{{\bf
9}}\right) &=&\frac{1}{2\Gamma^2}\left[ 2\epsilon
F^{2}L^{2}+1\right] \,, \label{KilNorm}
\end{eqnarray}
where we have defined
\[
E^{2}=\frac{1}{2F^{2}}\left[ 4F^{4}+2\left( 1+2F^{2}L^{2}\right)
^{2}+\left( 1-\frac{| A| }{|B|}\right) F^{2}L^{2}\right] \ .
\]
The vector $K_{{\bf 9}}$ is spacelike for $\epsilon=1,0$ and
time-like for large $ |t|$ in the cases $\left(
\epsilon,\alpha\right) =\left( -1,1\right) ,\left( -1,0\right) $.
Otherwise its causal character depends strongly on the actual
values of $ t$ and $y$. The vector $K_{{\bf 8}}$ is timelike in
the cases $ \left( \epsilon,\alpha\right) =\left( 0,1\right)
,\left( -1,1\right) ,\left( -1,0\right) $, while for $\left(
\epsilon,\alpha\right) =\left( 1,1\right) $ it becomes spacelike
for large $|t|$. For all other cases the causal character of $K_{
{\bf 8}}$ changes with $t$ and $y$.

The Killing algebra is given by
\begin{eqnarray}
\left[ K_{{\bf i}},K_{{\bf j}}\right]  &=&\varepsilon
_{ijk}K_{{\bf k}}\ , \nonumber \\ \left[ K_{{\bf 3+i}},K_{{\bf
3+j}}\right] &=&\epsilon\,\varepsilon _{ijk}K_{{\bf k} }\ ,
\nonumber \\ \left[ K_{{\bf i}},K_{{\bf 3+j}}\right]
&=&\varepsilon _{ijk}K_{{\bf 3+k}}\ , \nonumber \\ \left[ K_{{\bf
6+i}},K_{{\bf j}}\right] &=&0=\left[ K_{{\bf 6+i}},K_{{\bf
3+j}}\right] \ , \nonumber \\ \left[ K_{{\bf 7}},K_{{\bf
8}}\right]  &=&K_{{\bf 9}}\ , \nonumber
\\ \left[ K_{{\bf 8}},K_{{\bf 9}}\right]  &=&-\epsilon\alpha
K_{{\bf 7}}+\delta_{\alpha}{}^{0}K_{ {\bf 8}}\ ,  \nonumber \\
\left[ K_{{\bf 9}},K_{{\bf 7}}\right]  &=&-\alpha K_{{\bf
8}}+\delta_{\alpha}{}^{0}K_{ {\bf 7}}\ . \label{KilAlg}
\end{eqnarray}
The Killing algebra for the different admissible values of the
parameters $\epsilon$ and $\alpha$ is classified in Table~A1. The
Killing vectors $K_{{\bf 8,9}}$ have components in the bulk ($y$)
direction, while $K_{{\bf 1-7}}$ are confined to the $y=$const
sections, which is the reason we list their algebras separately in
Table~A1.

\begin{table}[tb]
\caption{Killing algebras for different values of $\epsilon$ and
$\alpha$} \vspace*{0.3cm}
\begin{tabular}{||c||c|c|c||}
\hline\hline $\epsilon$ & $1$ & $0$ & $-1$ \\
\hline$\alpha$ & $1$ & $1$ & $0\, , ~~\pm 1$ \\
\hline\hline $K_{{\bf 1-7}}$ & $so(4)\oplus {\bf {R}}$
    & $e(3)\oplus {\bf {R} }$ & $so(1,3)\oplus {\bf {R}}$ \\
\hline$K_{{\bf 1-9}}$ & $so(4)\oplus so(1,2)$ &
                           $e(3)\oplus e(1,1)$ &
                           $so(1,3)\oplus so(1,2)$
\\ \hline\hline
\end{tabular}
\end{table}


\newpage
\section*{References}

\begin{thebibliography}{99}
\bibitem{gen}
See, e.g., the recent reviews: M J Duff, hep-th/0012249; G T
Horowitz, hep-th/0011089; A Sen, hep-lat/0011073

\bibitem{RS} L Randall and R Sundrum 1999 {\em Phys. Rev. Lett.} {\bf 83}
4690

\bibitem{SMS}  T Shiromizu, K Maeda and M Sasaki 2000 {\it Phys. Rev. D}
{\bf 62} 024012

\bibitem{BDEL}  P Bin\'{e}truy, C Deffayet, U Ellwanger and D Langlois 2000
{\it Phys. Lett. B} {\bf 477} 285

\bibitem{Maartens}  R Maartens 2000 {\it Phys. Rev. D} {\bf 62}
084023; R Maartens, gr-qc/0101059

\bibitem{mwbh} R Maartens, D Wands, B A Bassett and I P C Heard
2000 {\it Phys. Rev. D} {\bf 62} 041301

\bibitem{MSM}  S Mukohyama, T Shiromizu and K Maeda 2000 {\it Phys. Rev. D}
{\bf 62\ }024028

\bibitem{BCG}  P Bowcock, C Charmousis and R Gregory 2000 {\it Class.
Quantum Grav. }{\bf 17} \ 4745

\bibitem{cs}
A Campos and C F Sopuerta 2001 {\em Phys. Rev. D} {\bf 63} 104012;
A Campos and C F Sopuerta 2001 {\em Phys. Rev. D} {\bf 64} 104011

\bibitem{Mannheim}  P D Mannheim, hep-th/0009065

\bibitem{Kraus}
P Kraus 1999 {\it JHEP }{\bf 12} 011; D Birmingham 1999 {\it
Class. Quantum Grav. }{\bf 16} 1197; R Emparan, C V Johnson and R
C Myers 1999 {\it Phys. Rev. D} {\bf 60\ }104001; R Emparan 1999
{\it JHEP }{\bf 06} 036


\end{thebibliography}
\end{document}